\documentclass[preprint,preprintnumbers,showpacs,aps,prd,amssymb]{revtex4-1}

\usepackage{graphicx}
\usepackage{bm}
\usepackage{amsmath}

\def\calH{{\cal H}}
\def\calL{{\cal L}}
\def\calO{{\cal O}}
\def\calU{{\cal U}}

\def\qbar{{\bar q}}
\def\lbar{{\bar \ell}}

\def\ccb{{c_{cb}}}
\def\cnul{{c_{\nu\ell}}}

\def\Ds{{D^{(*)}}}
\def\RDs{{R(D^{(*)})}}

\def\dU{{d_{\cal U}}}
\def\LU{{\Lambda_{\cal U}}}

\def\SM{{\rm SM}}
\def\nn{\nonumber}

\begin{document}
\title{Unparticle effects on $B\to D^{(*)}\tau\nu$}
\author{Jong-Phil Lee}
\email{jongphil7@gmail.com}
\affiliation{Sang-Huh College,
Konkuk University, Seoul 05029, Korea}

\begin{abstract}
We examine the possible unparticle effects on $\RDs$ associated with $B\to D^{(*)}\tau\nu$ decays 
by minimum-$\chi^2$ fitting.
Recent measurements from Belle and LHCb are included in this analysis.
While it is true that the new experimental results of $\RDs$ get closer to the standard model predictions,
there are still rooms for new physics and unparticles are also one possibility.
Our best-fit values are $R(D)=0.456$ and $R(D^*)=0.270$, 
which are still far from the standard model values by more than ($R(D)$) or almost ($R(D^*)$) $2\sigma$.
We also find that the unparticle effects are quite safe to render the branching ratio 
${\rm Br}(B_c\to\tau\nu)$ less than 10\%.

\end{abstract}
\pacs{}

\maketitle
%
The standard model (SM) is a very successful theory in particle physics.
But there must be new physics (NP) beyond the SM for many reasons. 
Flavor physics is a good testing ground for NP.
Recently, $B$ factories and LHCb observed an excess of the semileptonic $B$ decay, $B\to D^{(*)}\tau\nu$
\cite{BaBar1,BaBar_PRL,Belle1,Belle1607,Belle1703,Belle1709,LHCb1,LHCb2}.
The anomaly is encoded in a ratio of the branching ratios 
\begin{equation}
\RDs\equiv\frac{{\rm Br}(B\to D^{(*)}\tau\nu)}{{\rm Br}(B\to D^{(*)}\ell\nu)}~,
\end{equation}
where $\ell=e,\mu$.
The SM predicts that \cite{Na,Fajfer}
\begin{eqnarray}
\label{RSM}
R(D)_\SM&=&0.300\pm0.008~,\nn\\
R(D^*)_\SM&=&0.252\pm0.003~.
\end{eqnarray}
Both measurements from $BABAR$ and Belle are larger than the SM predictions, Eq.\ (\ref{RSM}). 
Results of $BABAR$ are quite different from the SM prediction at the $3.4\sigma$ level 
($R(D)$ by $2.0\sigma$ and $R(D^*)$ by $2.7\sigma$) \cite{BaBar1,BaBar_PRL}.
$BABAR$ also excluded the type-II two-Higgs-doublet model (2HDM) 
where a charged Higgs could contribute to $\RDs$
 at the 99.8\% confidence level.
The Belle results are between Eq.\ (\ref{RSM}) and the $BABAR$ measurements, 
and compatible with the type-II 2HDM.
Very recently, LHCb reported new measurements of $R(D^*)$ consistent with the SM \cite{LHCb2}.
Previous results from LHCb are larger than the SM predictions by $2.1\sigma$ \cite{LHCb1}.
In a recent analysis we showed that with an anomalous $\tau$ couplings, 
any types of 2HDM is as good as another to fit the $\RDs$ data \cite{JPL}.
\par
In this work we examine the unparticle effects on $\RDs$.
Unparticles are hypothetical things which behave like a fractional number of particles \cite{Georgi}.
In this scenario, a scale-invariant hidden sector at high energy couples to the SM particles weakly
at some high scale $\LU$.
The low-energy effective description of the scale-invariant sector is the unparticles.
Unparticle effects on $B$ physics have been studied in many ways
\cite{Geng,Mohanta,Lenz,Parry,JPL2,Li,Chen}.
As will be shown later, unparticles contribute quite differently from other NP particles.
We check the compatibility of the unparticle scenario by global $\chi^2$ fitting to $\RDs$.
%
%
\par
The relevant Lagrangian involving scalar unparticles $\calO_\calU$ coupled to the left-handed currents is given by
\begin{equation}
 \calL_\calU
=\frac{c_{q'q}}{\Lambda_\calU^\dU}\bar q'\gamma_\mu(1-\gamma_5)q~\partial^\mu\calO_\calU
  +\frac{c_{\ell'\ell}}{\Lambda_\calU^\dU}\bar\ell'\gamma_\mu(1-\gamma_5)\ell~\partial^\mu\calO_\calU~,
\label{LU}
\end{equation}
where $c_{q'q, \ell'\ell}$ are dimensionless couplings, and $\dU$ is the scaling dimension of $\calO_\calU$.
By the unitarity constraints, $\dU\ge 1$ \cite{unitarity}.
The effective Hamiltonian for $q\to q'\ell\ell'$ is then
\begin{eqnarray}
\label{Heff}
\calH_{eff}^\calU&=&
-\frac{A_\dU e^{-i\phi_\calU}}{2\sin\dU\pi}\frac{m_\ell c_{\ell'\ell}}{s^{2-\dU}\LU^{2\dU}}
\left[
    (\qbar'q)(\lbar'\ell)c_{q'q}(m_{q'}-m_q)
  +(\qbar'\gamma_5 q)(\lbar'\ell)c_{q'q}(-m_{q'}-m_q)\right.
\nn\\
&&\left.
  +(\qbar'q)(\lbar'\gamma_5\ell)c_{q'q}(-m_{q'}+m_q)
  +(\qbar'\gamma_5 q)(\lbar'\gamma_5\ell)c_{q'q}(m_{q'}+m_q)\right]~,
\end{eqnarray}
where
\begin{eqnarray}
\label{AdU}
A_\dU&=&\frac{16\pi^{5/2}}{(2\pi)^{2\dU}}
		\frac{\Gamma(\dU+1/2)}{\Gamma(\dU-1)\Gamma(2\dU)}~,\\
\label{phidU}		
\phi_\calU&=&(\dU-2)\pi~,
\end{eqnarray}
and $s\equiv (p_\ell+p_{\ell'})^2$.
%
%
\par
As discussed in \cite{Bardhan}, operators $\calO_{VL}$ and $\calO_{SL}$ contribute to $R(D)$
while operators $\calO_{VL}$, $\calO_{AL}$, and $\calO_{PL}$ do to $R(D^*)$, where
\begin{eqnarray}
\calO_{VL} &=& \left(\qbar'\gamma^\mu q\right)\left(\lbar'\gamma_\mu P_L\ell\right)~,~~~
\calO_{AL} = \left(\qbar'\gamma^\mu\gamma_5 q\right)\left(\lbar'\gamma_\mu P_L\ell\right)~,\nn\\
\calO_{SL} &=& \left(\qbar' q\right)\left(\lbar' P_L\ell\right)~,~~~
\calO_{PL} = \left(\qbar'\gamma_5 q\right)\left(\lbar' P_L\ell\right)~,
\end{eqnarray}
with $P_L=(1-\gamma_5)/2$.
While $\calO_{VL}$ affects both $R(D)$ and $R(D^*)$, $\calO_{VL}$ alone cannot provide 
satisfactory explanations for the experimental data \cite{Bardhan}. 
We expect the scalar unparticles can do the job and check the possibility.
\par
In this analysis we do not consider the vector unparticles.
Vector unparticles contribute through
\begin{equation}
\frac{c_V}{\LU^{d_V-1}}\qbar'\gamma_\mu(1-\gamma_5)q~\calO_\calU^\mu~,
\end{equation}
where $d_V$ is the scaling dimension of the vector unparticle operator $\calO_\calU^\mu$.
The unitarity constraint requires that $d_V\ge 3$ \cite{unitarity}.
Typically the contribution amounts to $\sim(m_B^2/\LU^2)^{d_V-1}$
while that of scalar unparticles is $\sim(m_B^2/\LU^2)^\dU$.
One can expect that effects of vector unparticles are very suppressed compared to 
those of scalar ones due to the unitarity constraints \cite{JPL2}.
%
%
\par
The decay rates of $B\to\Ds\ell\nu$ mediated by $\calO_\calU$ are given by
\begin{equation}
\Gamma^\Ds = \Gamma^\Ds_\SM +\Gamma^\Ds_{\rm mix} +\Gamma^\Ds_\calU~.
\end{equation}
The differential decay rates for $B\to D\ell\nu$ are given by
\begin{eqnarray}
\label{dGDdsSM}
\frac{d\Gamma^D_\SM}{ds}&=&
\frac{G_F^2 |V_{cb}|^2}{96\pi^3m_B^2}\left\{
 4m_B^2P_D^2\left(1+\frac{m_\ell^2}{2s}\right)|F_1|^2 \right. \\\nn
&&
\left. +m_B^4\left(1-\frac{m_D^2}{m_B^2}\right)^2\frac{3m_\ell^2}{2s}|F_0|^2\right\}
 \left(1-\frac{m_\ell^2}{s}\right)^2P_D~,
\\
\label{dGDdsmix}
\frac{d\Gamma^D_{\rm mix}}{ds}&=&
\frac{G_F}{\sqrt{2}}\frac{V_{cb}^*}{16\pi^3}
\left(\kappa_\calU\ccb\cnul\cos\phi_\calU\right)m_B^2m_\ell \\\nn
&&\times
\left(1-\frac{m_D^2}{m_B^2}\right) |F_0|^2
\left(1-\frac{m_\ell^2}{s}\right)^2 P_D~,
\\
\label{dGDdsU}
\frac{d\Gamma^D_\calU}{ds}&=&
\frac{m_B^2}{32\pi^3}
\left| \kappa_\calU\ccb\cnul \right|^2
|F_0|^2 s\left(1-\frac{m_\ell^2}{s}\right)^2 P_D~,
\end{eqnarray}
where 
\begin{equation}
\label{kappaU}
\kappa_\calU=
\frac{A_\dU}{2\sin\dU\pi}\frac{m_\ell}{s^{2-\dU}\LU^{2\dU}}~,
\end{equation}
is the unparticle factor and
\begin{equation}
\label{PD}
P_D\equiv
\frac{\sqrt{s^2+m_B^4+m_D^4-2(sm_B^2+sm_D^2+m_B^2m_D^2)}}{2m_B}~,
\end{equation}
is the momentum of $D$ in the $B$ rest frame.
The form factors $F_0$ and $F_1$ are given by
\begin{eqnarray}
\label{F01}
F_0 &=& \frac{\sqrt{m_Bm_D}}{m_B+m_D} (w+1)S_1~,\\
F_1 &=& \frac{\sqrt{m_Bm_D}(m_B+m_D)}{2m_BP_D}\sqrt{w^2-1}V_1~,
\end{eqnarray}
where
\begin{eqnarray}
V_1(w)&=&
V(1)\left[1-8\rho_D^2z(w)+(51\rho_D^2-10)z(w)^2-(252\rho_D^2-84)z(w)^3\right]~,\\
S_1(w)&=&
V_1(w)\left\{
1+\Delta\left[-0.019+0.041(w-1)-0.015(w-1)^2\right]\right\}~,
\end{eqnarray}
with
\begin{eqnarray}
w&=&\frac{m_B^2+m_D^2-s}{2m_Bm_D}~,~~~
z(w)=\frac{\sqrt{w+1}-\sqrt{2}}{\sqrt{w+1}+\sqrt{2}}~,\\
\rho_D^2&=&1.186\pm0.055~,~~~
\Delta=1\pm1~.
\end{eqnarray}
\par
For $B\to D^*\ell\nu$ decay,
\begin{eqnarray}
\label{dGDsdsSM}
\frac{d\Gamma^{D^*}_\SM}{ds} 
&=& 
\frac{G_F^2|V_{cb}|^2}{96\pi^3m_B^2}\left[
  \left( |H_+|^2+|H_-|^2+|H_0|^2\right)\left(1+\frac{m_\ell^2}{2s}\right)
  +\frac{3m_\ell^2}{2s}|H_s|^2\right] \\\nn
&&\times
  s\left(1-\frac{m_\ell^2}{s}\right)^2P_{D^*}~,\\
\label{dGDsdsmix}
\frac{d\Gamma^{D^*}_{\rm mix}}{ds} 
&=& 
\frac{G_F}{\sqrt{2}}\frac{V_{cb}^*}{4\pi^3}
\left( \kappa_\calU\ccb\cnul\cos\phi_\calU \right)m_\ell
A_0^2 \left(1-\frac{m_\ell^2}{s}\right)^2P_{D^*}^3~,\\
\label{dGDsdsU}
\frac{d\Gamma^{D^*}_\calU}{ds} 
&=&
\frac{1}{8\pi^3}
\left| \kappa_\calU\ccb\cnul \right|^2
A_0^2 s\left(1-\frac{m_\ell^2}{s}\right)^2P_{D^*}^3~,
\end{eqnarray}
where $P_{D^*}=P_D(m_D\to m_{D^*})$.
The form factors are given by
\begin{eqnarray}
\label{Hpm0s}
H_\pm(s) &=&
(m_B+m_{D^*})A_1(s)\mp\frac{2m_B}{m_B+m_{D^*}}P_{D^*} V(s)~,\\
H_0(s) &=&
\frac{-1}{2m_{D^*}\sqrt{s}}\left[
\frac{4m_B^2P_{D^*}^2}{m_B+m_{D^*}}A_2(s)
-(m_B^2-m_{D^*}^2-s)(m_B+m_{D^*})A_1(s)\right]~,\\
H_s(s) &=&
\frac{2m_B P_{D^*}}{\sqrt{s}} A_0(s)~,
\end{eqnarray}
where
\begin{eqnarray}
\label{A012V}
A_1(w^*) &=& \frac{w^*+1}{2}r_{D^*}h_{A_1}(w^*)~,\\
A_0(w^*) &=& \frac{R_0(w^*)}{r_{D^*}}h_{A_1}(w^*)~,\\
A_2(w^*) &=& \frac{R_2(w^*)}{r_{D^*}}h_{A_1}(w^*)~,\\
V(w^*)    &=& \frac{R_1(w^*)}{r_{D^*}}h_{A_1}(w^*)~,\\
\end{eqnarray}
with
\begin{equation}
\label{wsrDs}
w^* = \frac{m_B^2+m_{D^*}^2-s}{2m_Bm_{D^*}}~,~~~
r_{D^*} = \frac{2\sqrt{m_Bm_{D^*}}}{m_B+m_{D^*}}~,
\end{equation}
and
\begin{eqnarray}
\label{hR}
h_{A_1}(w^*)&=&
h_{A_1}(1)\left[1-8\rho_{D^*}^2z(w^*)+(53\rho_{D^*}^2-15)z(w^*)^2
-(231\rho_D{^*}^2-91)z(w^*)^3\right]    ~,\\
R_0(w^*) &=&
R_0(1)-0.11(w^*-1)+0.01(w^*-1)^2    ~,\\
R_1(w^*) &=&    
R_1(1)-0.12(w^*-1)+0.05(w^*-1)^2    ~,\\
R_2(w^*) &=&    
R_2(1)+0.11(w^*-1)-0.01(w^*-1)^2     ~.
\end{eqnarray}
Here\cite{Dhargyal}
\begin{eqnarray}
\rho_{D^*}^2 &=& 1.207\pm0.028~,~~~ R_0(1)=1.14\pm 0.07~,\\
R_1(1)&=&1.401\pm0.033~,~~~ R_2(1)=0.854\pm 0.020~.
\end{eqnarray}
%
\par
The experimental data for our fits are given in Table \ref{T1}.
The $R(D^*)$ values from the Belle get slightly closer to the SM predictions.
%
\begin{table}
\begin{tabular}{|c|| cc |}\hline
              & ~$R(D)$ & ~$R(D^*)$  \\\hline\hline
 BABAR & ~$0.440\pm0.058\pm0.042$ & ~$0.332\pm0.024\pm0.018$ \cite{BaBar1} \\
 Belle(2015) & ~$0.375\pm0.064\pm0.026$ & ~$0.293\pm0.038\pm0.015$ \cite{Belle1} \\
 Belle(2016) & ~$-$ & ~$0.302\pm0.030\pm0.011$ \cite{Belle1607} \\
 Belle(1703) & ~$-$ & ~$0.276\pm0.034^{+0.029}_{-0.026}$ \cite{Belle1703} \\
 Belle(1709) & ~$-$ & ~$0.270\pm0.035^{+0.028}_{-0.025}$ \cite{Belle1709} \\
 LHCb(1506) & ~$-$ & ~$0.336\pm0.027\pm0.030$ \cite{LHCb1} \\
 LHCb(1711) & ~$-$ & ~$0.286\pm0.019\pm0.025\pm0.021$ \cite{LHCb2} \\\hline
 \end{tabular}
\caption{Experimental data for $R(\Ds)$.
The uncertainties are $\pm$(statistical)$\pm$(systematic).
For the third uncertainty of LHCb(1711), see \cite{LHCb2} for details.}
\label{T1}
\end{table}
%
As can be seen in Eqs.\ (\ref{dGDdsmix}), (\ref{dGDdsU}), (\ref{dGDsdsmix}), (\ref{dGDsdsU}), 
the unparticle couplings appear only in the form of $\ccb\cnul$.
In our analysis $\ccb\cnul$, $\dU$, and $\LU$ are fitting parameters to minimize $\chi^2$,
which is defined by
\begin{equation}
\chi^2=\sum_i\frac{(x_i-\mu_i)^2}{(\delta\mu_i)^2}~,
\end{equation}
where the $x_i$'s are model predictions and the $(\mu_i\pm\delta\mu_i)$'s are experimental data.
New physics effects in $\RDs$ could affect the $B_c\to\tau\nu$ decay \cite{Grinstein,Akeroyd}.
Scalar unparticles contribute to the branching ratio of $B_c\to\tau\nu$ as
\begin{equation}
\label{BrBc}
{\rm Br}(B_c\to\tau\nu)={\rm Br}(B_c\to\tau\nu)_\SM\left| 1+r_\calU\right|^2~,
\end{equation}
where
\begin{eqnarray}
\label{BrBcSM}
{\rm Br}(B_c\to\tau\nu)_\SM&=&
\tau_{B_c}\frac{G_F^2|V_{cb}|^2}{8\pi}m_{B_c}m_\tau^2f_{B_c}^2
\left(1-\frac{m_\tau^2}{m_{B_c}^2}\right)^2~,\\
\label{rU}
r_\calU&=&
\frac{\ccb\cnul}{\sqrt{2}G_Fm_{B_c}^2V_{cb}}
\frac{A_\dU e^{-i\phi_\calU}}{\sin\dU\pi}
\left(\frac{m_{B_c}}{\LU}\right)^{2\dU}~.
\end{eqnarray}
Here $\tau_{B_c}$ and $f_{B_c}$ are the lifetime and the decay constant of $B_c$, respectively.
\par
\begin{figure}
\includegraphics[scale=0.2]{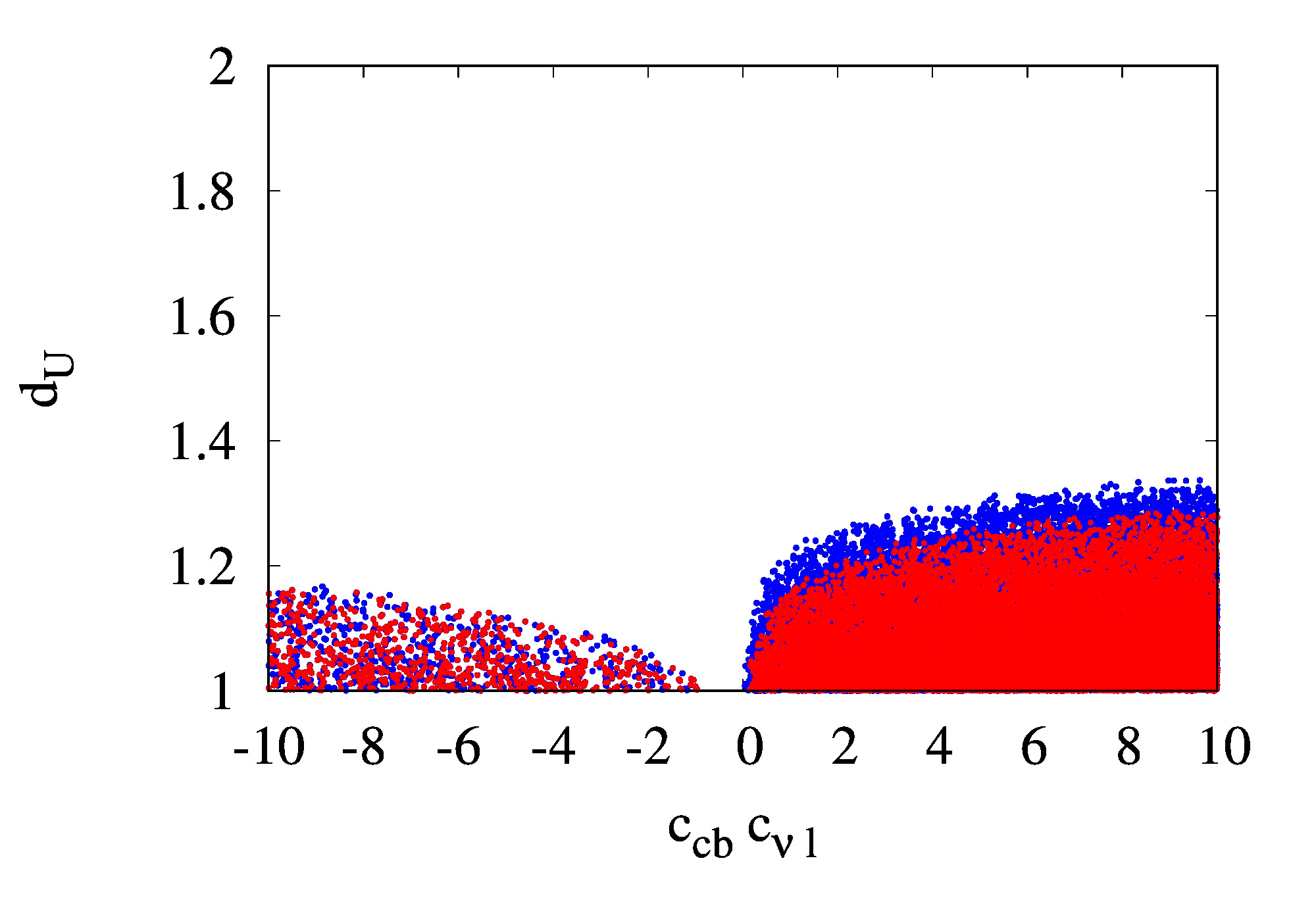}
\caption{\label{Fig1} Allowed region for $\ccb\times\cnul$ vs $\dU$ at the $1\sigma$ (red) 
and the $2\sigma$ (blue) levels.}
\end{figure}
\begin{figure}
\includegraphics[scale=0.2]{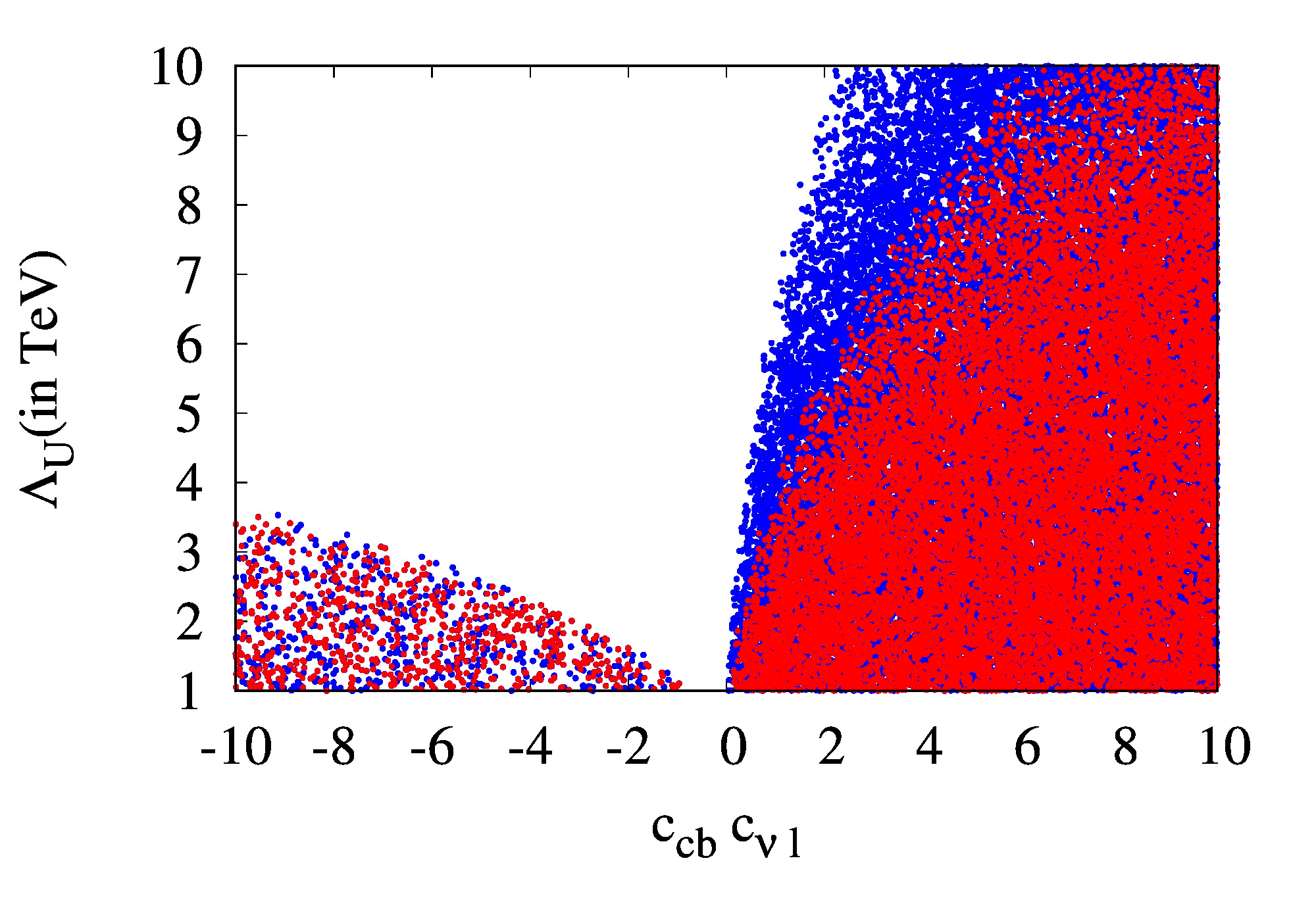}
\caption{\label{FigLUc} Allowed region for $\ccb\cnul$ vs $\LU$ at the $1\sigma$ (red) 
and the $2\sigma$ (blue) levels.
$\LU$ is scanned over $1\le\LU\le 10$ TeV.}
\end{figure}
\begin{figure}
\includegraphics[scale=0.2]{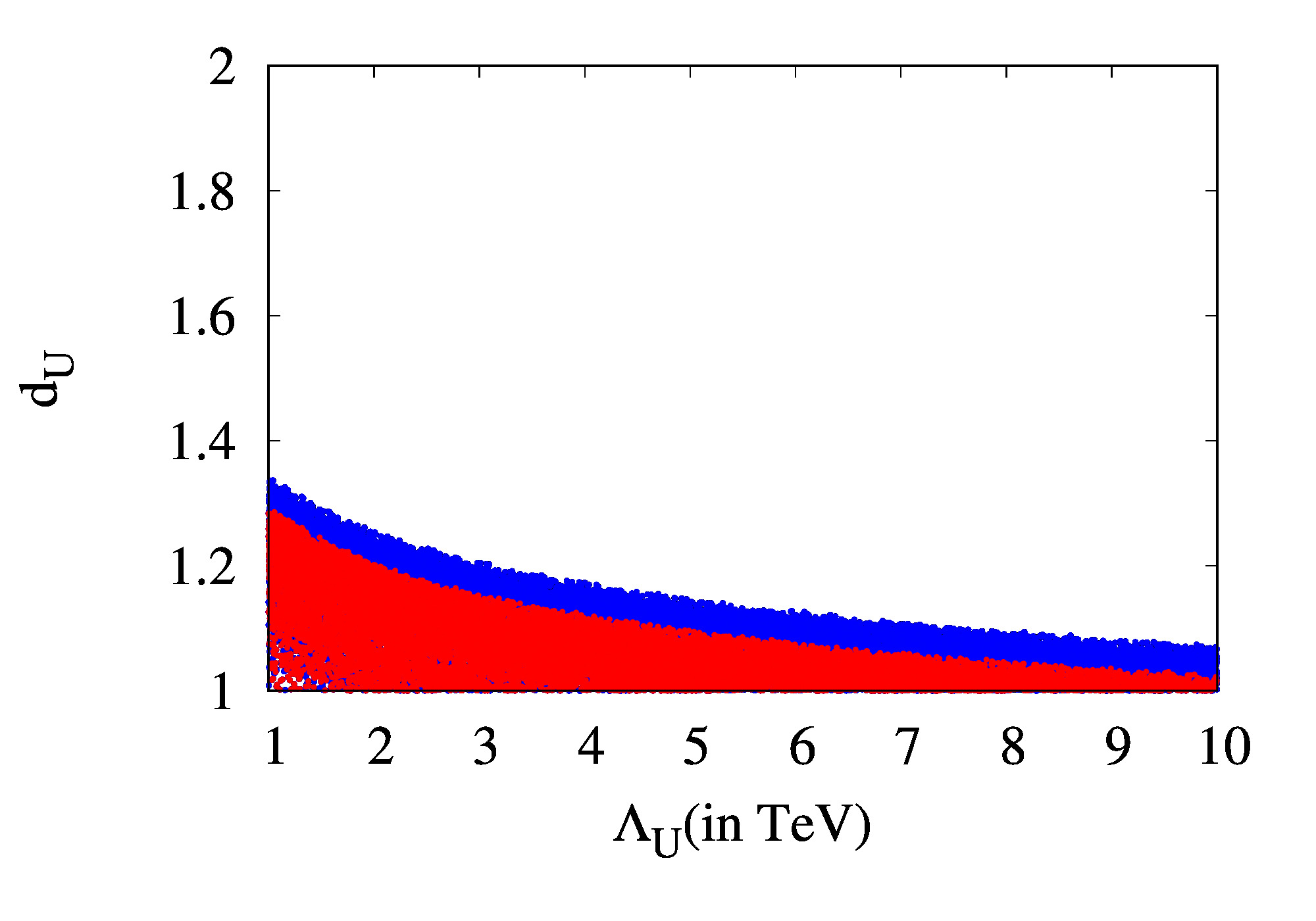}
\caption{\label{FigdULU} Allowed region for $\LU$ vs $\dU$ at the $1\sigma$ (red) 
and the $2\sigma$ (blue) levels.
$\LU$ is scanned over $1\le\LU\le 10$ TeV.}
\end{figure}
\begin{figure}
\begin{tabular}{cc}
\includegraphics[scale=0.13]{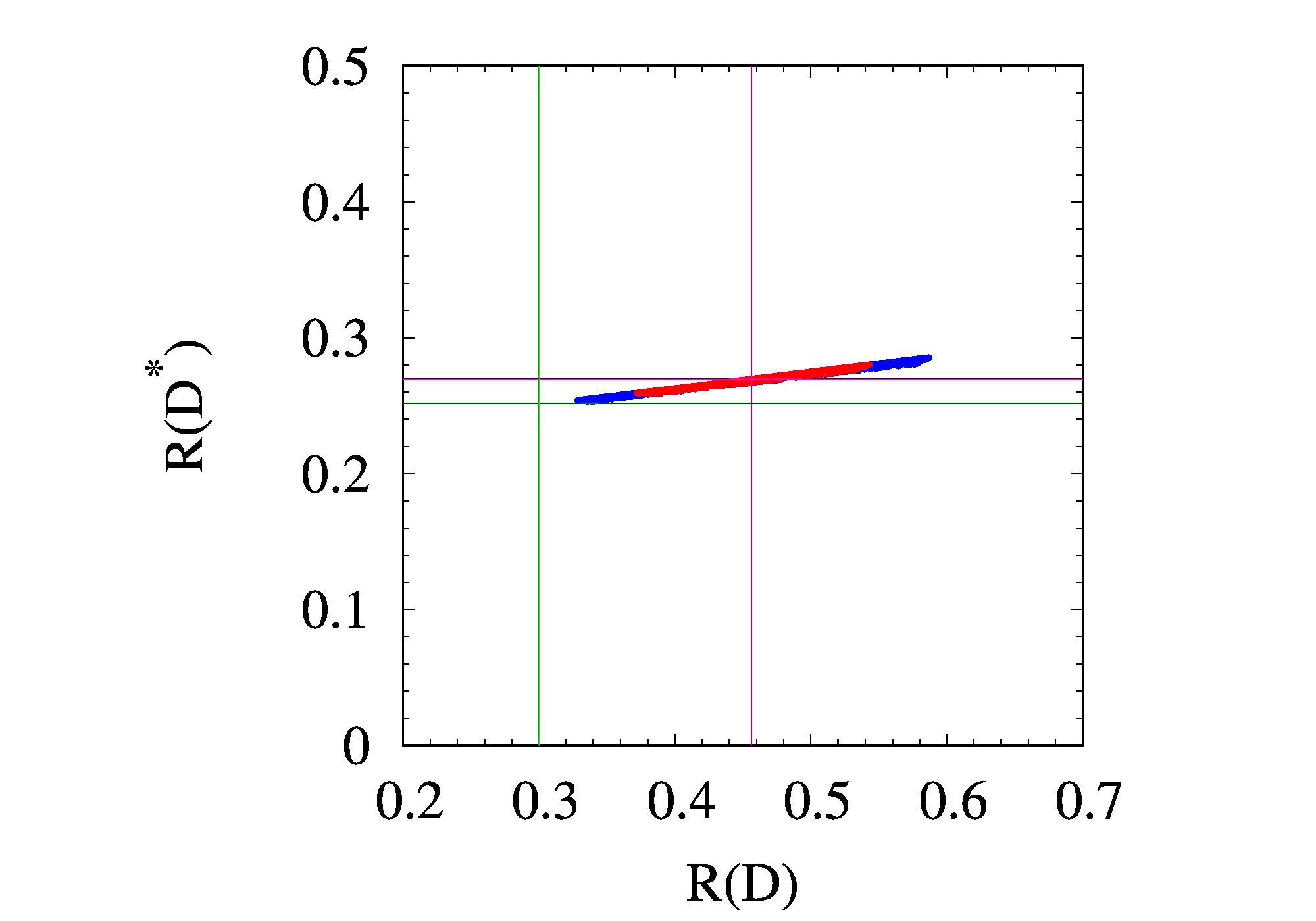}&
\includegraphics[scale=0.13]{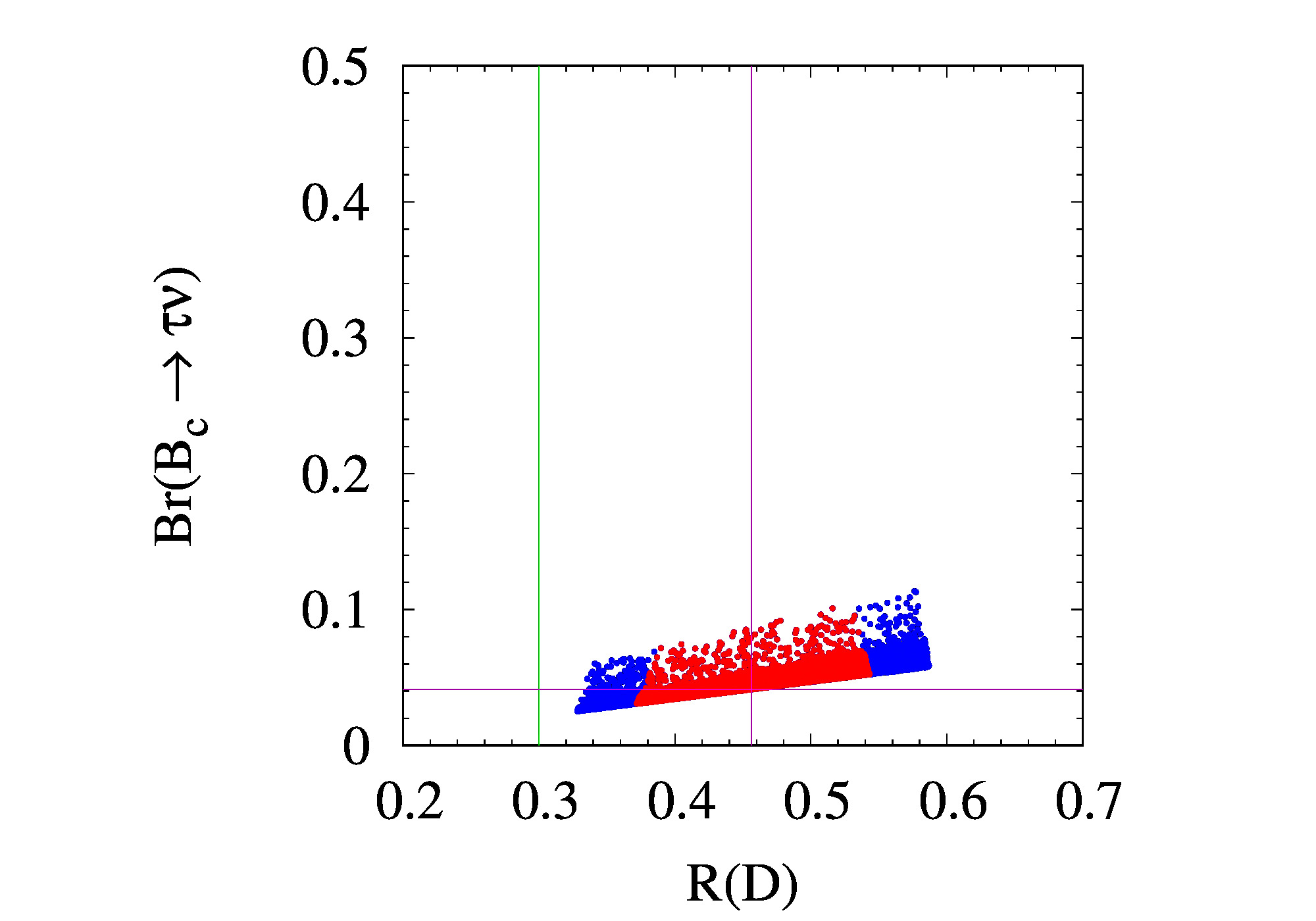}\\
(a)&(b)
\end{tabular}
\caption{\label{FigRD} Allowed region in the $R(D)$-$R(D^*)$ plane (panel (a))
and the $R(D)$-${\rm Br}(B_c\to\tau\nu)$ plane (panel (b))
at the $1\sigma$ (red) and the $2\sigma$ (blue) levels.
Green lines are the SM predictions while magenta ones are the best-fit points.}
\end{figure}

Figure \ref{Fig1} shows the allowed region in the $\ccb\cnul$-$\dU$ plane at the
$1\sigma$ (red) and the $2\sigma$ (blue) levels.
Note that the unparticle contribution comes as
\begin{equation}
\label{mixU}
\frac{d\Gamma^\Ds_{\rm mix}}{ds}+\frac{d\Gamma^\Ds_\calU}{ds}
\sim
\ccb\cnul\left(\frac{s}{\LU^2}\right)^\dU
+\left| \ccb\cnul\left(\frac{s}{\LU^2}\right)^\dU \right|^2~.
\end{equation}
Here $s=(p_\tau+p_\nu)^2\le(m_B-m_{D^(*)})^2$ while $\LU\sim\calO({\rm TeV})$, 
thus suppression of $(s/\LU^2)$ gets stronger as $\dU$ gets larger.
Thus for small values of $| \ccb\cnul |$, large $\dU$ is not allowed because in this case
the unparticle contribution becomes very small.
%
In Fig.\ \ref{FigLUc} $\ccb\cnul$ and $\LU$ are shown at the $1\sigma$ (red) and $2\sigma$ (blue) levels.
Usually new physics (NP) effects appear as
\begin{equation}
\label{NP}
\sim \lambda_{\rm NP}^\alpha
\left(\frac{M_{\rm EW}}{M_{\rm NP}}\right)^\beta~,
\end{equation}
where $\lambda_{\rm NP}$ is a new coupling, $M_{\rm EW}$ is the electroweak scale, and
$M_{\rm NP}$ is the NP scale, with some fixed powers of $\alpha$ and $\beta$.
Typically at lowest order $\alpha=\beta=2$.
In this case for large values of $\lambda_{\rm NP}$, small values of $M_{\rm NP}$ are not allowed
because the value of Eq.\ (\ref{NP}) could be very large.
%
%
In the unparticle scenario as in this analysis, $\beta=2\dU$ is a model parameter which varies freely.
The result is that the suppression of Eq.\ (\ref{NP}) for large $\lambda_{\rm NP}$ is possible
because $(M_{\rm EW}/M_{\rm NP})^{2\dU}$ can be small enough for large $\dU$.
This feature is shown in Fig.\ \ref{FigdULU}.
Figure \ref{FigdULU} shows the allowed region in the $\LU$-$\dU$ plane.
Note that for small values of $\LU$ around $\sim1$ TeV (and for large $\ccb\cnul$) 
large values of $\dU$ are allowed. 
On the other hand, the NP effects of Eq.\ (\ref{NP}) should not be too small to account for 
anomalies well beyond the SM predictions.
Roughly speaking, $\beta$ and $M_{\rm NP}$ can not be large simultaneously.
As expected from Eq.\ (\ref{mixU}),  for large values of $\LU$ only small values of $\dU$ are permitted.
%
%
In Fig.\ \ref{FigRD} allowed ranges of $R(D)$ vs $R(D^*)$ (panel (a)) and 
$R(D)$ vs ${\rm Br}(B_c\to\tau\nu)$ (panel (b)) are shown.
The SM predictions of $R(D)$ and $R(D^*)$ are slightly off the $2\sigma$ region.  
As shown in Fig.\ \ref{FigRD}(b), the branching ratio of $B_c\to\tau\nu$ is mostly below $\sim10\%$.
Our result for ${\rm Br}(B_c\to\tau\nu)$ is safe enough to satisfy a stronger constraint of \cite{Akeroyd}
where the branching ratio should not exceed 10\%.
%
%
The best-fit values are summarized in Table \ref{T2}.
Note that $\chi^2_{\rm min}/{\rm d.o.f}$ is not far from 1,
thus it can be said that unparticles fit the data well.
%
\begin{table}
\begin{tabular}{ccccccc }\hline\hline
 $R(D)$ & $R(D^*)$ & ~${\rm Br}(B_c\to\tau\nu)$ & $\dU$ & $\ccb\cnul$ & ~$\LU$ (in TeV)   & ~$\chi^2_{\rm min}/{\rm d.o.f}$ \\ \hline
 $0.456$	& $0.270$ & $4.134\times 10^{-2}$ & $1.000$ & ~$2.306$ & $4.326$  & $1.671$ \\\hline\hline
 \end{tabular}
\caption{The best-fit values. 
$\chi^2_{\rm min}/{\rm d.o.f}$ is the minimum value of $\chi^2$ per degree of freedom.}
\label{T2}
\end{table}
\par
Recently the CMS collaboration announced the lower limits of $\LU$ with respect to $\dU$
at high energy collisions \cite{CMS13TeV}.
According to the Fig.\ 10 of \cite{CMS13TeV}, $\LU$ must be larger than $\sim 10^2$ TeV 
for $\dU\lesssim 1.4$, and the lower limit of $\LU$ decreases for larger $\dU$ until
$\LU\gtrsim 0.3$ TeV for $\dU=2.2$.
The results are for the fixed coupling, $\lambda=1$.
In our language, $\lambda = \left[(m_j-m_k)/\LU\right]c_{jk}$.
Thus $\lambda=1$ corresponds to very large values of $c_{jk}\gtrsim 10^3$.
For such a large $c_{jk}$, $\LU$ must be large enough to keep the unparticle contribution moderate,
as can be seen in Eq.\ (\ref{mixU}).
For example, if $c_{cb}c_{\nu\ell}\sim 10^6$ then $\LU^2\gtrsim 10^6$ for $\dU\sim 1$
because $\LU\sim$ (a few) TeV for $c_{cb}c_{\nu\ell}\sim 1$ in Fig.\ \ref{FigLUc}.
In this reason, our results are compatible with recent LHC data above TeV scale.
%
\par
In conclusion, we have investigated the unparticle contributions to $\RDs$.
We only considered scalar unparticles because contributions from vector unparticles are expected to be very small.
We fit the data by minimizing $\chi^2$ and found that $\chi^2_{\rm min}/{\rm d.o.f}$ is smaller than that of
our previous work with 2HDM ($\sim 2.9$) \cite{JPL}.
At lowest order scalar unparticles contribute to $\RDs$ as $\sim \ccb\cnul(s/\LU^2)^\dU$.
New physics scale $\LU$ could be around $\sim 1{\rm TeV}$ thanks to $(s/\LU^2)^\dU$ suppression.
Our best-fit values of $R(D)=0.456$ and $R(D^*)=0.270$ are larger than the SM predictions 
by almost ($R(D^*)$) or more than ($R(D)$) $2\sigma$, as shown in Fig.\ \ref{FigRD}.
It is well known that the NP effects for $\RDs$ would also affect $B_c\to\tau\nu$ decay,
and ${\rm Br}(B_c\to\tau\nu)$ could provide a strong constraint for NP.
We found that scalar unparticles can render the branching ratio less than 10\% (Fig.\ \ref{FigRD} (b)).
More data for $\RDs$ and ${\rm Br}(B_c\to\tau\nu)$ would check the plausibility of the unparticle scenario.
%

\end{document}